# Emerging optical spin and chirality by 3D evanescent field coupling

Liang Fang[1,2*] and Jian Wang[2*]

[1]*School of Physics and Electronics, Hunan University, Changsha 410082, Hunan, China*

[2]*Wuhan National Laboratory for Optoelectronics and School of Optical and Electronic Information, Huazhong University of Science and Technology, Wuhan 430074, Hubei, China*

**Corresponding author: liangfang@hnu.edu.cn, jwang@hust.edu.cn.*

Optical spin and chirality play key roles in engineering photonic emission and light-matter interactions. Here we show that 3D evanescent coupling of guided modes by strongly confined waveguides can extrinsically produce optical spin and chirality. These emerging optical quantities are essentially derived from an intrinsic phase retardation of π/2 between two coupled modes. For a directional coupler with 2×2 input ports, the produced optical spin and chirality are inherently characterized by the side-locked and path-locked phenomena, respectively. We numerically demonstrate that these optical spin and chirality can be ubiquitously produced by higher-order mode coupling and other structural waveguides. These new quantities in optical waveguides may pave the way to exploit near-field photonic spin and on-chip chiral light-matter interactions.

## I. INTRODUCTION

Polarization, spin, and chirality are three fundamental quantities of optical fields that have greatly advanced the fundamentals and applications of light [1-8]. Optical spin or spin angular momentum (SAM) originates from the rotation of electric and magnetic vectors in circularly or elliptically polarized electromagnetic fields. Optical chirality is associated with the helicity of light determined by the degree of circular polarization [9,10]. These two quantities play key roles in various light-matter interactions and chirality-related optical effects, such as optical rotation effect, circular dichroism (CD), and Raman optical activity [11–15]. Especially, optical chirality can be largely enhanced by electromagnetic resonances via plasmonic nanostructures or dielectric metamaterials, which may open new opportunities for chiral analysis and separation [16-20].

In a circularly polarized plane wave [see Fig. 1(a)], the electric and magnetic fields contribute equally to longitudinal optical spin, and the optical chirality can be naturally produced [21,22]. However, when strongly confined by high-index contrast optical waveguides, the transverse-magnetic (TM) and transverse-electric (TE) modes always have different propagation constants (speed) due to the non-degenerated modes [23-25]. As a result, these quasi-linearly polarized waveguide modes in the strongly confined conditions usually can not form longitudinal spin and optical chirality. It is worth noting that the weakly guided optical fibers with circular symmetry definitely support the circularly polarized modes due to the degenerated even and odd components [24,25], thus can form optical spin and chirality.

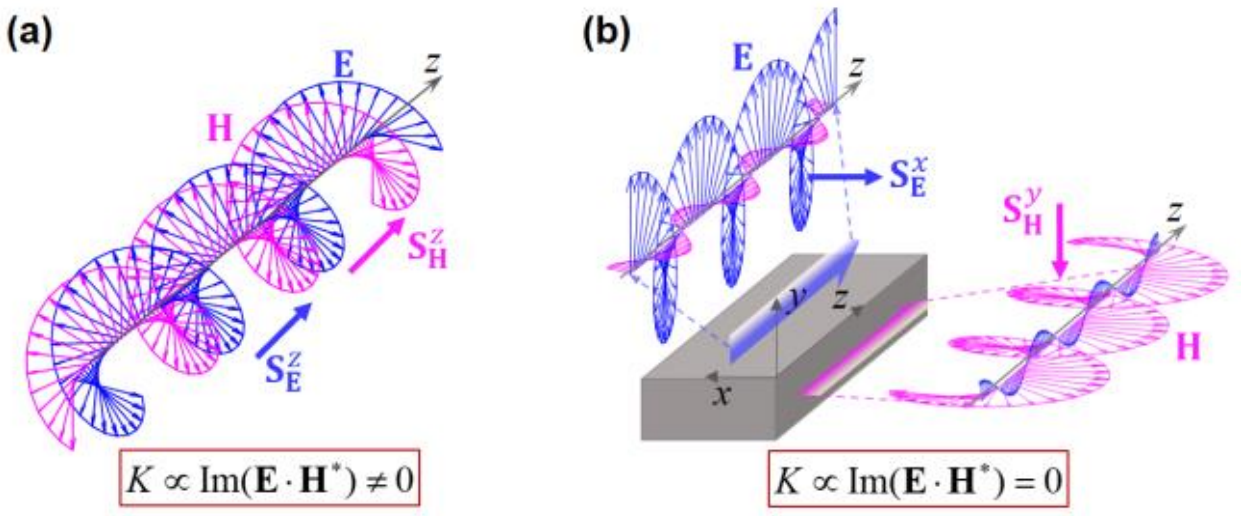

FIG. 1. Sketch of optical spin produced in different conditions. (a) Circularly polarized electromagnetic fields produce optical spin and chirality. (b) Transverse and longitudinal field components with a phase retardation of π/2 produce transverse spin but no optical chirality in high-index contrast waveguides. Here **E** and **H** stand for the electric and magnetic field, respectively. **S** and *K* indicate optical spin and chirality, respectively. The blue and magenta arrows show the electric and magnetic polarization (or spin) vectors, respectively.

Despite not producing longitudinal spin and optical chirality, high-index contrast waveguides usually possess extraordinary transverse spin, which is determined by the intrinsic phase retardation of π/2 between the transverse and longitudinal field components [26,27], [see Fig. 1(b)]. This transverse spin ubiquitously exists in strongly confined conditions, such as the high-index contrast waveguides and strongly focused situations [28-39]. In these cases, the electric and magnetic transverse spin components are non-overlapped because of the broken equivalence between electric and magnetic field components [31,36]. Such transverse spin with an intrinsic spin-momentum locking effect, can be used for engineering unidirectional photon emission in chiral photon circuits [40-44]. Furthermore, this transverse spin within structured light fields has also been witnessed to produce chiral-dependent lateral forces that enable all-optical chiral separation [45–47].

In recent decades, structured light, known as the electromagnetic fields with tailorable or shapeable degrees of freedom (DoFs) in terms of intensity, phase, polarization, and angular momentum, has become one of the hottest research topics [48-50]. Waveguide modes, naturally tailored (constrained) by the boundaries of optical waveguides, inherently exhibit spatial (non-uniform) structural characteristics [24,25,51-53]. It is known that waveguide modes can easily produce mutual coupling under index perturbation, governed by coupled-mode theory [25]. However, almost all previous studies on coupling focused on the management of optical power, wavelengths, polarization, and modes [25,54-58], few attentions are paid to the near-field manifestations of coupled modes.

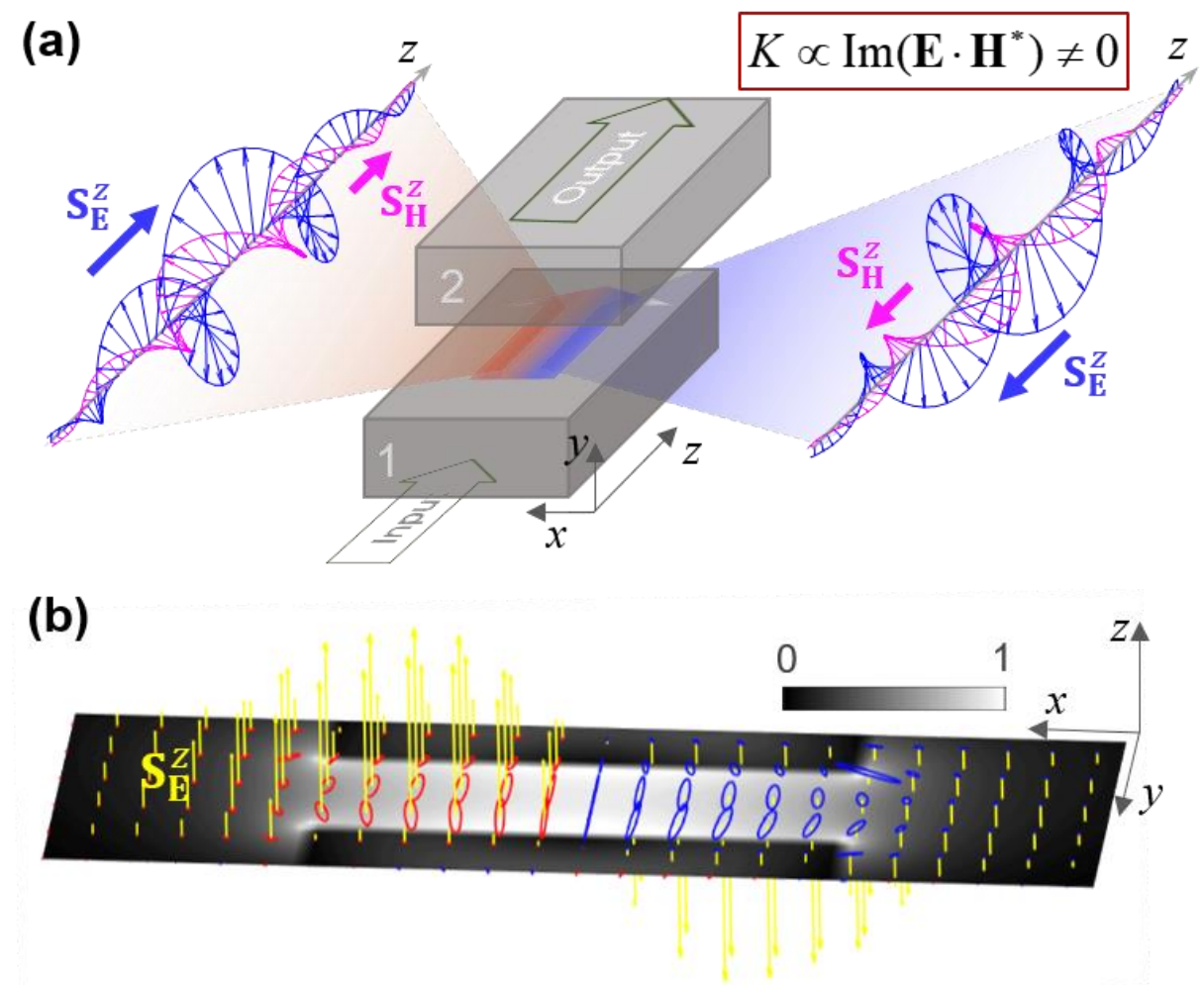

FIG. 2. Optical spin and chirality produced by mode coupling. (a) Sketch of directional coupler producing the electric and magnetic polarization ellipses, and the non-zero optical chirality. (b) Electric polarization ellipses and spin (yellow arrows) with odd-symmetric distributions in the gap. The red and blue ellipses indicate the right- and left-handed polarization states, respectively. The gray map shows the normalized field intensity in the *x*-*y* plane.

Here we investigate the near-field spin and optical chirality of directionally coupled modes with spatial vectorial structures. Remarkably, the near-field polarization helicity and longitudinal spin can produce in the gap of coupled waveguides due to the intrinsic phase retardation of π/2 between coupled modes [see Fig. 2]. This can be analogous to a quarter-wave plate inducing a phase retardation of π/2 between two orthogonal polarizations along the slow and fast axes, respectively. Despite not completely equivalent electric and magnetic components, such electromagnetic helicity can still create optical chirality. In this paper, we thoroughly demonstrate the optical properties of the produced longitudinal spin and chirality by directional coupling between waveguides.

## II. THEORETICAL RESULTS

The directionally coupled modes between two adjacent waveguides follow coupled-mode theory [25]. For simplification here, we assume that the coupled modes completely satisfy the phase-matching condition. Accordingly, the superposed field $\tilde{\boldsymbol{\psi}}$ with coupled modes in the coupling region can be respectively written as,

$$\tilde{\boldsymbol{\psi}}_1(z) = [A\cos(\kappa z)\mathbf{E}_a + iB\sin(\kappa z)\mathbf{E}_a]\cdot e^{i\beta z}, \quad (1)$$

$$\tilde{\boldsymbol{\psi}}_2(z) = [iA\sin(\kappa z)\mathbf{E}_b + B\cos(\kappa z)\mathbf{E}_b]\cdot e^{i\beta z}, \quad (2)$$

where $\mathbf{E}$ indicates the spatial electric field distribution of waveguide eigenmodes, the subscript $a$ or $b$ denotes the mode supported in waveguide 1 or 2, respectively. In this case, the propagation constant is $\beta_a = \beta_b = \beta = n_{eff}k$, $n_{eff}$ is the modal effective index, and $k = 2\pi/\lambda$ is the wave vector. The coupling coefficient $\kappa$ refers to an overlap integral between two modes [25]. $A$ and $B$ determine the incident conditions of coupling system.

To investigate near-field spin and chirality of the coupled electromagnetic fields, a superposed electric field should be considered, i.e., $\tilde{\mathbf{E}} = \tilde{\boldsymbol{\psi}}_1 + \tilde{\boldsymbol{\psi}}_2$, as well as the magnetic field component $\tilde{\mathbf{H}} = -i\nabla\times\tilde{\mathbf{E}}/\omega\mu_0$. Optical spin is generally given by $\mathbf{S} = \mathrm{Im}(\varepsilon_0\tilde{\mathbf{E}}^*\times\tilde{\mathbf{E}}/4\omega\mu_b + \mu_0\tilde{\mathbf{H}}^*\times\tilde{\mathbf{H}}/4\omega\varepsilon_b)$, where $\varepsilon_0$ and $\mu_0$ are the vacuum permittivity and permeability, respectively, $\varepsilon_b$ and $\mu_b$ are the relative permittivity and permeability of background, respectively, and $\omega$ is the angular frequency of monochromatic light. Here focusing on the incidence condition ($A$=1 and $B$=0), from Eqs. (1) and (2), the electric spin produced by evanescent coupling can be given as,

$$S_{\mathrm{E}}^z(\mathbf{r},z) = \frac{\varepsilon_0}{4\omega\mu_b}\sin(2\kappa z)\Phi(\mathbf{r}), \quad (3)$$

where $\Phi(\mathbf{r}) = E_x^a E_y^b - E_x^b E_y^a$. Such spin, parallel to the coupling direction, originates from the intrinsic phase retardation of π/2 between two coupled modes. The corresponding magnetic spin is

$$S_{\mathrm{H}}^z(\mathbf{r},z) = \frac{\sin(2\kappa z)}{4\omega^3\varepsilon_b\mu_0}(\beta^2-\kappa^2)[\Phi(\mathbf{r})+\xi(\mathbf{r})], \quad (4)$$

and the resulting optical chirality,

$$K(\mathbf{r},z) = \varepsilon_0\varepsilon_b\mu_b\beta\sin(2\kappa z)[\Phi(\mathbf{r})+\gamma(\mathbf{r})], \quad (5)$$

where $c$ is the light speed in vacuum, and $\xi(\mathbf{r})$ and $\gamma(\mathbf{r})$ are correction terms for the spatial distributions of magnetic spin and optical chirality, respectively, see the Appendix A. These correction factors $\xi(\mathbf{r})$ and $\gamma(\mathbf{r})$ are caused by the considerable longitudinal component $E_z$ of waveguide modes, breaking the symmetry of the electric and magnetic field distributions. By contrast, the circularly polarized light remains this symmetry between the electric and magnetic components [see Fig. 1(a)], where the longitudinal field is $E_z \approx 0$, giving rise to $\xi(\mathbf{r}) \approx 0$ and $\gamma(\mathbf{r}) \approx 0$. Note that if the coupling length ($L$) satisfies $\kappa L = (1+2n)\pi/4$ with $n$=0,1,2,3…, corresponding to the 3dB positions where modes are equally coupled into two waveguides, the produced spin and chirality reach the maximum values.

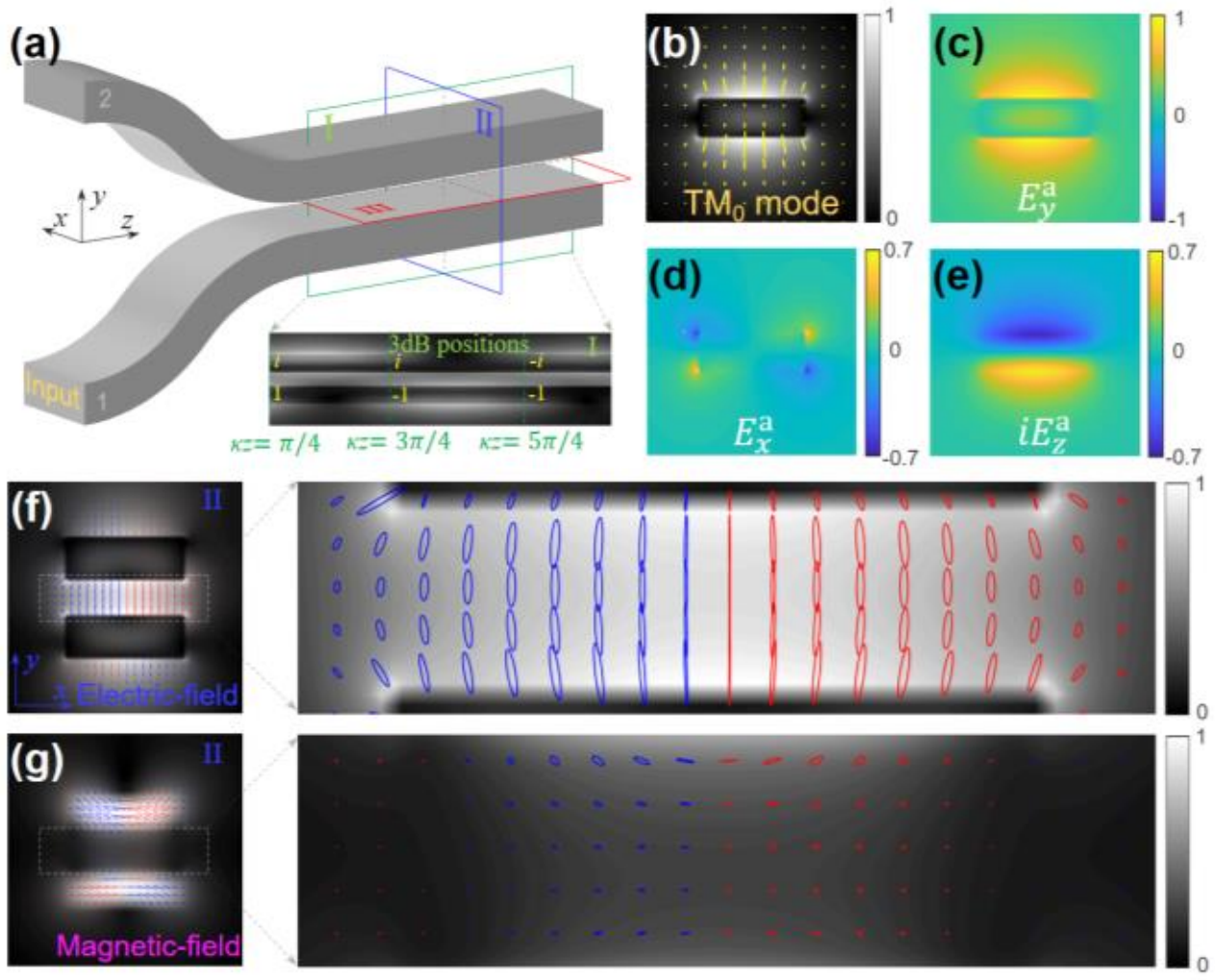

FIG. 3. Field and polarization evolution by evanescent coupling between silicon-based waveguides. (a) Sketch of a directional coupler. The inset shows periodic coupling with phase changes monitored in the plane I. (b) $TM_0$ mode with electric field and polarization distributions. (c)-(e) Three polarization components of $TM_0$ mode. (f) and (g) Near-field electric and magnetic polarization ellipse distributions in one 3dB position, respectively. The insets show the enlarged polarization states, the red and blue ellipses indicate the left- and right-handed elliptical polarizations, respectively.

## III. NUMERICAL RESULTS AND ANALYSES

In this section, we present the numerical results of the near-field coupled spin and chirality produced by mode coupling between two identical silicon-based waveguides 1 and 2 [see Fig. 3(a)]. The two coupled modes $a$ and $b$, respectively supported by waveguides in Eqs. (1) and (2), naturally satisfy the full phase-matching condition. Here the incident mode is $TM_0$ mode from waveguide 1. It is worth noting that when strongly guided by two-dimensional (2D) strip waveguides here, the TM modes do not feature purely vertical polarization $E_y^a$ along the $y$-axis at all, [see Figs. 3(b)-3(d)]. The field component $E_x^a$ with odd-symmetric distribution becomes remarkable and thus responsible for the production of coupled spin and chirality of evanescent fields in the gap of waveguides.

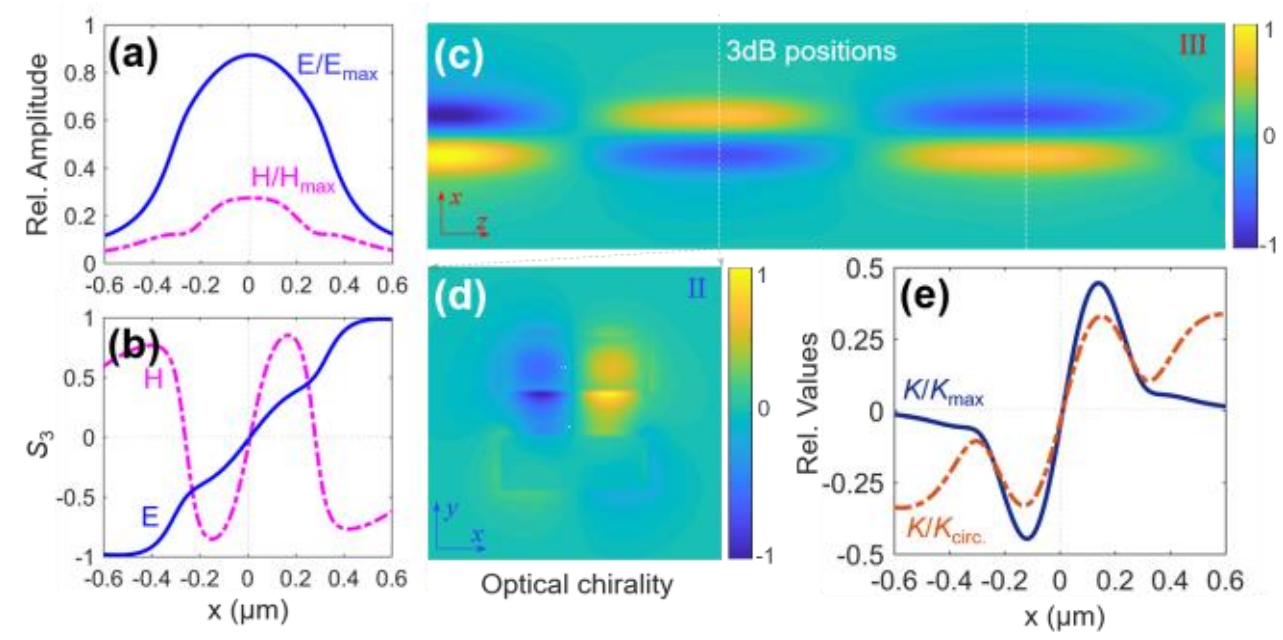

FIG. 4. Numerical results of optical spin and chirality produced by coupled waveguides. (a) Electric and magnetic field amplitudes calculated in the center line of gap along the $x$-axis, relative to their maximum values in Figs. 3(f) and 3(g), respectively. (b) The calculated Stokes parameter $S_3$ of electric and magnetic polarization ellipses. (c) Distribution of optical chirality calculated in the monitor plane III, shown in Fig. 3(a). (d) Distribution of optical chirality in the same plane with Figs. 3(f) and 3(g). (e) Optical chirality along the center line of gap in Fig. 4(d), calculated as the ratio of this quantity to the maximum value (mazarine line) and that corresponding to the circularly polarized field with the same electric field amplitude (brown dotted line), respectively.

Here the directional coupler consists of silicon-based waveguides, the waveguide has the thickness of 0.22 μm, the width of 0.6 μm, and the gap between two waveguides is 0.18 μm. The numerical data were obtained by means of the finite-different time-domain solutions. The periodic coupling along the propagation direction is gotten from the plane I [see the inset of Fig. 3(a)]. The coupled modes between two waveguides have the phase relationship determined by Eqs. (1) and (2). We numerically calculate the electric and magnetic fields with polarization distributions in the plane II, located near a 3dB position where modes are coupled with the same power into two waveguides, as shown in Figs. 3(f) and 3(g), respectively. Both the electric and magnetic polarization ellipses show odd-symmetric distributions along the $x$-axis, so do the resulting spin, according well with Eqs. (3) and (4). The relative electric and magnetic field amplitudes along the center line of gap are shown in Fig. 4(a). We calculated the average quantum number of SAM per photon, quantified by Stokes parameter $S_3$ for both electric and magnetic components [see Fig. 4(b)] [59]. This is essentially determined by the odd-symmetric distributions of $x$-polarized electric components [see Fig. 3(d)]. The out-of-step slopes of electric and magnetic curves result from the large longitudinal component of guided modes, associated with the correction term $\xi(\mathbf{r})$ in Eq. (4) that breaks the electromagnetic symmetric distributions.

The produced optical chirality [see Figs. 4(c)], along with optical spin, periodically changes their signs along the coupling direction. To investigate optical chirality produced by the coupled waveguides, we calculated two kinds of relative optical chirality [see the results in Fig. 4(e)]. One is calculated as the ratio of this quantity to the maximum chirality in the cross section [Fig. 4(d)]. The other is to the circularly polarized field with the same electric field amplitude in the gap, defined as $K_{\text{circ.}} = \omega\varepsilon_0\varepsilon_b\mu_b |\tilde{\mathbf{E}}|^2 / c$. This relative value $K / K_{\text{circ.}}$ cannot reach 1, due to the broken symmetry between the electric and magnetic components, essentially arising from the large longitudinal component of guided modes as well. Note that even though the coupled waveguides produce optical spin and chirality, their integral values are zero in the volume of coupling regions, i.e., $\int S_{\mathrm{E}}^{z} dV = 0$, $\int S_{\mathrm{H}}^{z} dV = 0$, and $\int K dV = 0$.

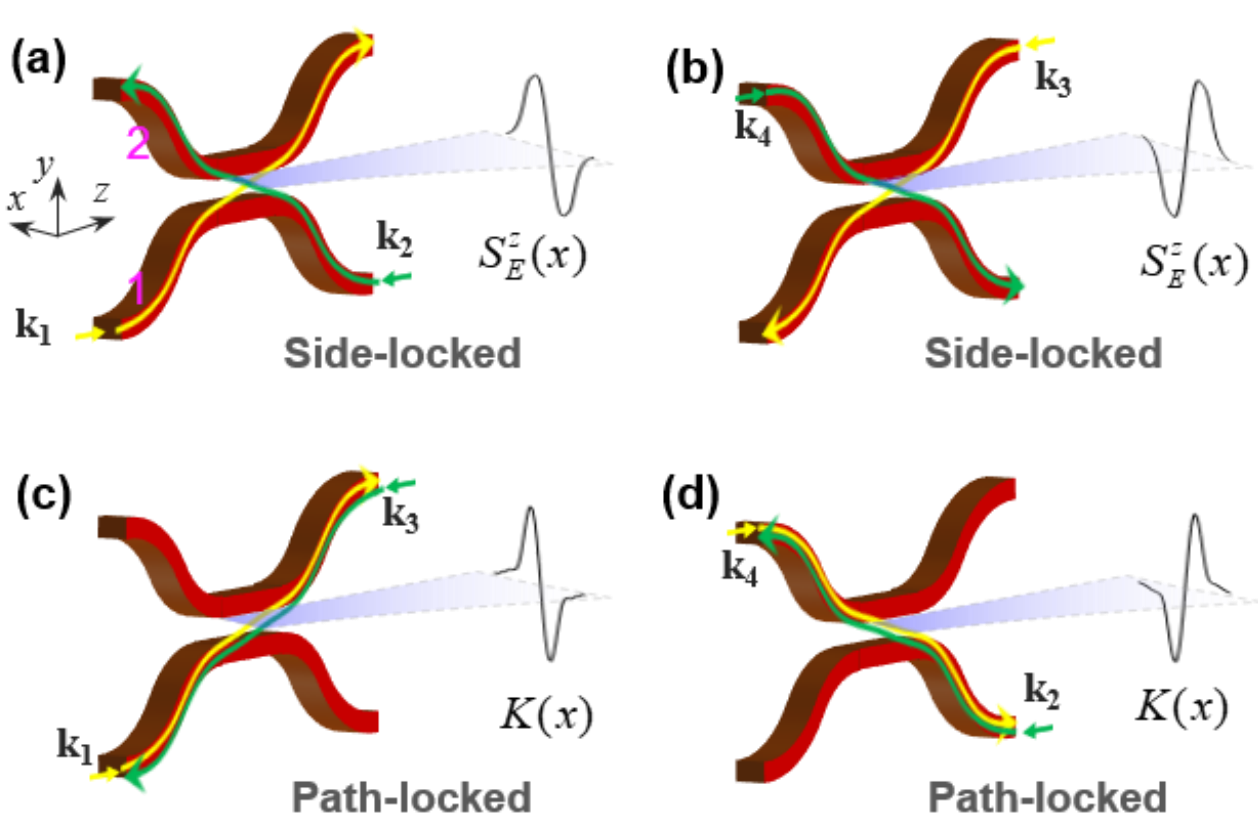

FIG. 5. Sketch of side- and path-locked effects for two quantities produced in a directional coupling system. (a) and (b) Side-locked optical spin produced under modal incidences from different waveguide sides. (c) and (d) Path-locked optical chirality produced under modal incidences giving different optical paths. Note that the optical spin and chirality distributions along the *x*-axis are depicted in the center of gap.

We investigate the distributions of coupled spin and optical chirality under different incident conditions of $\mathrm{TM}_0$ mode from the different imports of a $2\times2$ directional coupler. There exist two locking phenomena [see Fig. 5]. One is the side-locked optical spin that exhibits vectorial distributions being locked to the sides of the coupled waveguides 1 and 2 [see Figs. 5(a) and 5(b)]. This can be visually explained by $C_{2y}$ rotational operation of the coupling system for the spin pseudovector. The other is the path-locked optical chirality with spatial distribution locked to the coupling path regardless of the coupling directions [see Figs. 5(c) and 5(d)]. This can be associated with the analogous effect of coupling system under time reversal (T) operation because of the T-even optical chirality [9,10,60]. If the input waveguide modes are counter-propagated, such T-even optical chirality may produce interference to form periodical constructive distributions in the coupling region, similar to the standing wave, which enables the potential to chiral separation for chiral nanoparticles [61]. If the input modes are co-propagated, for example, simultaneous inputs from $k_1$ and $k_4$, there are two situations to be considered and discussed. One is that when the complex amplitudes A and B in Eqs. (1) and (2) satisfy the relationship of A=±B, no spin and chirality can be produced, which can be explained by the destructive interference in Fig. 5. The other is the case of A=±iB, where the produced spin and chirality in the 3dB positions shift towards the propagation direction [see the inset of Fig. 1(a) and Fig. 4(c)]. All these can be theoretically inferred from Eqs. (1) and (2). We also investigate the evolution of the produced optical spin and chirality when changing the incident wavelengths and waveguide's thicknesses, respectively [see Fig. 6]. These comparison results demonstrate that the waveguide modes being closer to the cut-off condition can produce larger spin and chirality, because of the enhanced evanescent fields in the gap [compare Figs. 6(c) with 6(a) or 6(f) with 6(d)].

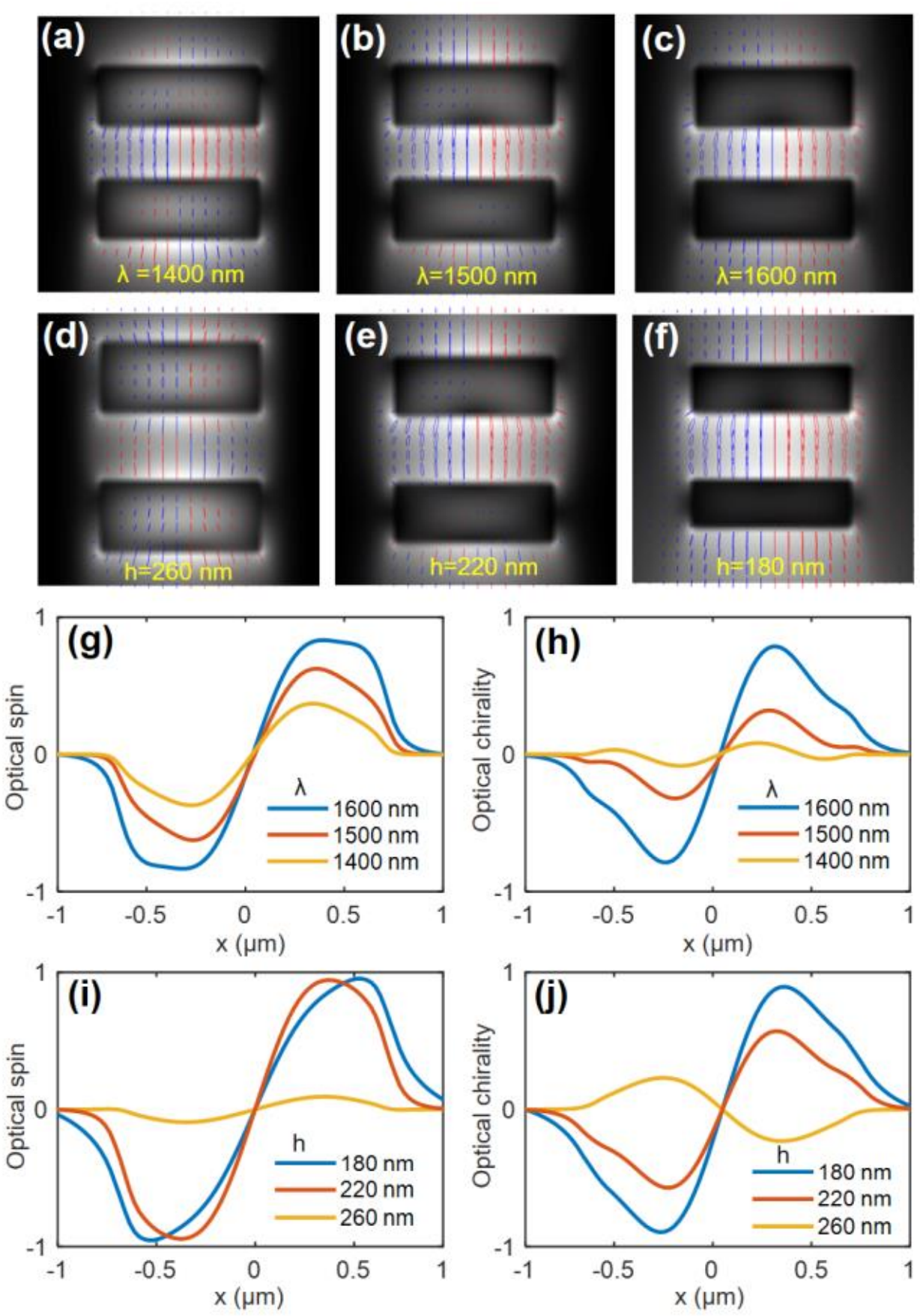

FIG. 6. Optical spin and chirality produced by coupled waveguides with different thicknesses or at different incident wavelengths. Near-field electric intensity and polarization ellipse distributions in the waveguide gap at the wavelength of

(a) 1400 nm, (b) 1500 nm, and (c) 1600 nm, respectively. In this case, the gap distance is set as 180 nm, and the waveguide thickness is fixed as 220 nm. Those distributions when the waveguide thickness is (d) 260 nm, (e) 220 nm, and (f) 180 nm, respectively. In this case, the gap distance is 240 nm, and the incident wavelength is fixed as 1550 nm. (g) Optical spin and (h) chirality produced along the center line of gap for (a)-(c), respectively. (i) and (j) Those values for (d)-(f). Note that all the relative values are calculated under the same input power.

## IV. COUPLED SPIN AND CHIRALITY BY OTHER WAVGUIDE MODES

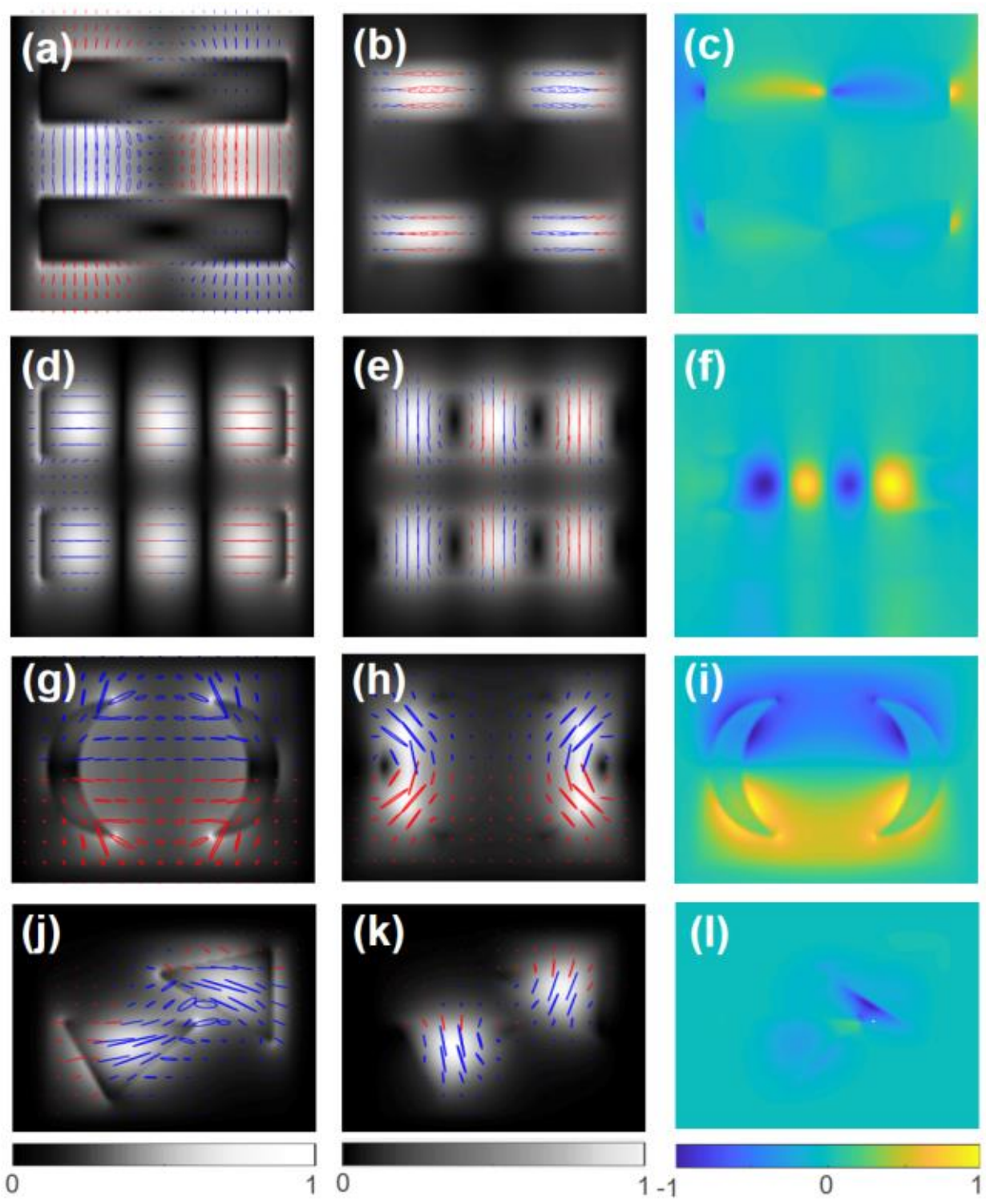

FIG. 7. Optical spin and chirality ubiquitously produced by evanescent coupling. Electric, magnetic spin and optical chirality produced by evanescent coupling of (a)-(c) higher-order $TM_1$ mode, (d)-(f) higher-order $TE_2$ mode in strip waveguides, fundamental modes in (g)-(i) crescent and (j)-(i) arbitrary triangle waveguides, respectively. (a), (d), (g), and (j) Electric field distributions with polarization states. (b), (e), (h), and (k) Magnetic field distributions with polarization states. (c), (f), (i), and (l) Distributions of the produced optical chirality.

Optical spin and chirality can be ubiquitously produced in arbitrary coupling systems of 3D structural modes supported by high-index contrast waveguides, beyond the degenerated modes in weakly guiding conditions, for example, fiber-guided linearly polarized (LP) modes. In this section, we show these two optical quantities produced by higher-order guided modes of strip waveguides and other arbitrary waveguide modes [see Fig. 7]. The multi-node evanescent fields of higher-order modes may form more complex spin and chirality distributions in the gap [see Figs. 7(a)-7(f)]. Without considering practical fabrication, the crescent and arbitrary triangle waveguides with silicon core in air are numerically simulated, and the resulting optical spin and chirality in the 3dB coupling positions are exhibited in Figs. 7(g)-7(l). These sufficiently demonstrate that the additional near-field spin component can be extrinsically created by the evanescent coupling of vectorial modes with 3D spatial polarization distributions. In contrast, the weakly guided LP modes and nearly pure TM and TE modes in one-dimensional slab waveguides [23-25,51] can hardly generate these spin and chirality, because of the uniform polarization components in the coupling region. The produced spin and chirality are characterized by symmetric spatial distributions in the cross section of coupled waveguides [see Figs. 6 and 7(a)-7(i)], which is determined by the symmetric field distributions due to the spatially symmetric waveguides. Especially, this symmetry of spin and chirality distributions can be broken by asymmetric coupled waveguides [see Figs. 7(j)-7(l)]. All these results show the universality of optical spin and chirality that can be produced by the evanescent of structural modes in the systems of optical waveguides.

## V. CONCLUSIONS

Electric and magnetic spin densities of circularly polarized electromagnetic fields share the same spatial distributions, because of the symmetry between electric and magnetic components. Accordingly, the produced optical chirality has been widely exploited for chiral light-matter interactions. However, in the system of high-index contrast waveguides, despite the available transverse spin, there is no optical chirality because of the non-overlapped electromagnetic fields. Since the modes in strongly confined waveguides can hardly produce optical chirality, the chiral light-matter interaction would be greatly limited on the waveguide-based photonic platforms. In this paper, we systematically investigate the optical spin and chirality extrinsically produced by the evanescent coupling of silicon-based waveguides. Similar to the spin-momentum locking transverse spin, the spatial distributions of optical spin and chirality in the directional coupler with $2\times2$ ports exhibit intrinsic side-locked and path-locked phenomena respectively. The gradient optical chirality with distributions possess the potential to chiral separation by the resulting chiral gradient forces [61]. These extraordinary spin and chirality induced by evanescent coupling may open new opportunities to develop waveguide optics, on-chip chiral separation and light-matter interactions.

## ACKNOWLEDGMENTS

This work was supported by the National Natural Science Foundation of China (NSFC) (62275092, 61905081), the Fundamental Research Funds for the Central Universities.

## APPENDIX A: THEORETICAL DERIVATION OF OPTICAL SPIN AND CHIRALITY BY MODE COUPLING

In this section, we focus on the detailed derivation of coupled electric and magnetic spin, as well as the resulting optical chirality, under the input case ($A$=1 and $B$=0). The electric field of the coupling system is

$$\tilde{\mathbf{E}}=[\cos(\kappa z)\mathbf{E}_a + j\sin(\kappa z)\mathbf{E}_b]\mathrm{e}^{i\beta z}, \tag{A1}$$

where the electric field of two coupled modes are $\mathbf{E}_a = E_x^a\mathbf{e}_x + E_y^a\mathbf{e}_y + E_z^a\mathbf{e}_z$, and $\mathbf{E}_b = E_x^b\mathbf{e}_x + E_y^b\mathbf{e}_y + E_z^b\mathbf{e}_z$. Generally, the transverse field components $E_x$ and $E_y$ of waveguide eigenmodes usually have a uniform phase, but an intrinsic phase retardation of π/2 relative to the longitudinal component ($E_z$). The magnetic field can be derived as

$$\tilde{\mathbf{H}} = -i\nabla\times\tilde{\mathbf{E}}/\omega\mu_0 = (H_x\mathbf{e}_x + H_y\mathbf{e}_y + H_z\mathbf{e}_z)e^{i\beta z}, \tag{A2}$$

where

$$\begin{aligned}
H_x &= \frac{-i}{\omega\mu_0}\{\cos(\kappa z)\frac{\partial E_z^a}{\partial y} + i\sin(\kappa z)\frac{\partial E_z^b}{\partial y} \\
&\quad -[\kappa\sin(\kappa z) + i\beta\cos(\kappa z)]E_y^a + [i\kappa\cos(\kappa z) + \beta\sin(\kappa z)]E_y^b\} \\
H_y &= \frac{i}{\omega\mu_0}\{\cos(\kappa z)\frac{\partial E_z^a}{\partial x} + i\sin(\kappa z)\frac{\partial E_z^b}{\partial x} \\
&\quad -[\kappa\sin(\kappa z) + i\beta\cos(\kappa z)]E_x^a + [i\kappa\cos(\kappa z) + \beta\sin(\kappa z)]E_x^b\} \\
H_z &= \frac{-i}{\omega\mu_0}[\cos(\kappa z)(\frac{\partial E_y^a}{\partial x} - \frac{\partial E_x^a}{\partial y}) + i\sin(\kappa z)(\frac{\partial E_y^b}{\partial x} - \frac{\partial E_x^b}{\partial y})]
\end{aligned} \quad . \tag{A3}$$

From Eq. (A3), ignoring the transverse spin, the density of electric spin produced by mode coupling is

$$\mathbf{S}_\mathrm{E}(\mathbf{r},z) = \frac{\varepsilon_0}{8\omega\mu_b}\sin(2\kappa z)\,\mathrm{Re}(\mathbf{E}_a\times\mathbf{E}_b^* + \mathbf{E}_a^*\times\mathbf{E}_b). \tag{A4}$$

Due to the intrinsic phase retardation of π/2 between the transverse and longitudinal field components, the spin produced by coupling must be longitudinal, as follows,

$$S_\mathrm{E}^z(\mathbf{r},z) = \frac{\varepsilon_0}{4\omega\mu_b}\sin(2\kappa z)\Phi(\mathbf{r}), \tag{A5}$$

where $\Phi(\mathbf{r}) = E_x^a E_y^b - E_x^b E_y^a$. The corresponding magnetic spin can be derived as

$$\begin{aligned}
S_\mathrm{H}^z(\mathbf{r},z) &= \frac{\mu_0}{4\omega\varepsilon_b}\mathrm{Im}(H_x^* H_y - H_x H_y^*) \\
&= \frac{\sin(2\kappa z)}{4\omega^3\varepsilon_b\mu_0}(\beta^2 - \kappa^2)[\Phi(\mathbf{r}) + \xi(\mathbf{r})]
\end{aligned} \quad , \tag{A6}$$

where the correction term $\xi(\mathbf{r})$ is

$$\begin{aligned}
\xi(\mathbf{r}) = \mathrm{Im}[&i\frac{\partial E_z^a}{\partial y}\frac{\partial E_z^b}{\partial x} - i\frac{\partial E_z^a}{\partial x}\frac{\partial E_z^b}{\partial y} - \kappa\frac{\partial E_z^a}{\partial y}E_x^a + \beta\frac{\partial E_z^a}{\partial y}E_x^b \\
&-\beta\frac{\partial E_z^b}{\partial y}E_x^a + \kappa\frac{\partial E_z^b}{\partial y}E_x^b + \kappa\frac{\partial E_z^a}{\partial x}E_y^a + \beta\frac{\partial E_z^b}{\partial x}E_y^a \\
&-\beta\frac{\partial E_z^a}{\partial x}E_y^b - \kappa\frac{\partial E_z^b}{\partial x}E_y^b]
\end{aligned} \quad , \tag{A7}$$

The optical chirality can be given by

$$K(\mathbf{r},z) = \frac{\omega\varepsilon_b\mu_b}{c^2}\mathrm{Im}(\tilde{\mathbf{E}}\cdot\tilde{\mathbf{H}}^*) = \varepsilon_0\varepsilon_b\mu_b\beta\sin(2\kappa z)[\Phi(\mathbf{r}) + \gamma(\mathbf{r})] \quad , \tag{A8}$$

with the correction term $\gamma(\mathbf{r})$

$$\begin{aligned}
\gamma(\mathbf{r}) = \mathrm{Re}[&\frac{i}{2}(E_x^a\frac{\partial E_z^b}{\partial y} - E_x^b\frac{\partial E_z^a}{\partial y}) - \frac{i}{2}(E_y^a\frac{\partial E_z^b}{\partial x} - E_y^b\frac{\partial E_z^a}{\partial x}) \\
&-\frac{i}{2}E_z^a(\frac{\partial E_y^b}{\partial x} - \frac{\partial E_x^b}{\partial y}) + \frac{i}{2}E_z^b(\frac{\partial E_y^a}{\partial x} - \frac{\partial E_x^a}{\partial y})]
\end{aligned} \quad , \tag{A9}$$